\documentclass[aps,twocolumn,showpacs]{revtex4}
\usepackage{graphicx}

\begin{document}

\title{Precision measurement of the environmental temperature by tunable
double optomechanically induced transparency with a squeezed field}
\author{Qiong Wang$^{1,2}$}
\author{Jian-Qi Zhang$^{1}$}
\email{Changjianqi@gmail.com}
\author{Peng-Cheng Ma$^{1,3}$}
\author{Chun-Mei Yao$^{2}$}
\author{Mang Feng$^{1}$}
\email{mangfeng@wipm.ac.cn}
\affiliation{$^{1}$ State Key Laboratory of Magnetic Resonance and Atomic and Molecular
Physics, Wuhan Institute of Physics and Mathematics, Chinese Academy of
Sciences, Wuhan 430071, China}
\affiliation{$^{2}$ College of Physics and Electronics, Hunan University of Arts and
Science, Changde 415000, China}
\affiliation{$^{3}$ School of Physics and Electronic Electrical Engineering, Huaiyin
Normal University, Huaian, 223300, China}

\begin{abstract}
A tunable double optomechanically induced transparency (OMIT) with a
squeezed field is investigated in a system consisting of an optomechanical
cavity coupled to a charged nanomechanical resonator via Coulomb
interaction. Such a double OMIT can be achieved by adjusting the strength of
the Coulomb interaction, and observed even with a single-photon squeezed
field at finite temperature. Since it is robust against the cavity decay,
but very sensitive to some parameters, such as the environmental
temperature, the model under our consideration can be applied as a quantum
thermometer for precision measurement of the environmental temperature
within the reach of current techniques.
\end{abstract}

\maketitle

\section{Introduction}

Electromagnetically induced transparency (EIT) is a kind of effect with a
narrow transparency window within an absorption line of atoms \cite%
{rmp-77-633}, due to quantum interference between two quantum pathways in $%
\Lambda$-type atoms. This effect plays a key role in modern quantum optics
experiments and applications, such as enhanced nonlinear susceptibility \cite%
{pra-81-053829}, optical switch \cite{pra-74-063831}, slow and fast lights
\cite{nature-397-594}, quantum memory \cite{Nat.Photonics-3-706,Laser
Photonics Rev-6-333,prl-110-213605}, quantum interference \cite%
{pra-90-023814} and vibrational cooling \cite{oe-21-29695}. Recently, the
study of the EIT has been extended to multi-channels, e.g., the double EIT
\cite{pra-83-053834,pra-84-033803}, and focused on simulation of other
physical phenomena, including Anderson localization \cite{pra-83-053847} and
quasi-charged particles \cite{pra-77-043813}.

The EIT analog occurring in an optomechanical system is called
optomechanically induced transparency (OMIT) \cite{science-330-1520}, which
is caused by quantum interference between two quantum channels in a $\Lambda$%
-type hybrid level configuration composed of photon states of the cavity and
phonon states of the optomechanical resonator \cite%
{pra-81-041803,pra-83-023823}. The OMIT has been explored both theoretically
\cite{pra-86-013815} and experimentally \cite{science-330-1520,
naturephoton-4-236,nature-471-204,nature-472-69}. Similar to the EIT, the
study of the originally defined OMIT \cite{pra-86-053806} was extended to
the double OMIT by two coupled optomechanical resonators \cite{pra-88-053813}
or by an optomechanical resonator coupled to other systems \cite%
{pra-90-023817,pra-90-043825}. Besides, the double OMIT was explored from
the fixed double OMIT \cite{pra-88-053813,jpb-47-055504,pra-90-023817} to
the tunable one involving a controllable coupling \cite{pra-90-043825}.

The present paper intends to investigate the unique behavior of a double
OMIT with a squeezed field. Due to involvement of the squeezed field, the
OMIT is robust against quantum noise and thus possible to be a candidate of
quantum memory \cite{pra-83-043826}. But if the model is extended to be a
double OMIT in a tunable manner, the physics turns to be largely different.
The key point is that the double OMIT is robust to the cavity decay and
quantum noise of the environment can be correlated to the
temperature-dependent noise. As such, we may carry out precision measurement
of the environmental temperature, assisted by the squeezed field and the
homodyne spectroscopy. Using other unique characteristics, precision
measurements of other parameters of the system are also available. This
implies that the model under our consideration is by no means a simple
extension of the previously considered OMIT \cite{pra-83-043826}, but with
much different characteristics and applications.

The temperature dependence is from the quantum field involved in our double
OMIT, by which we are able to know the environmental temperature through detecting the noise spectra of the
optomechanics \cite{pra-89-032114}. This is very different from the OMIT with classical lights
\cite{science-330-1520, pra-90-043825,pra-88-013804}, whose noise spectra have no relevance to the environmental
temperature even under the cryogenic condition \cite{science-330-1520}. In this context, our scheme is also
very different from the previous OMIT measurements resulted from the
properties of the OMIT spectra \cite{pra-86-053806,pra-90-043825}.
As a result, our scheme provides a new paradigm for precision
measurement based on the noise \cite{prl-106-140502,prl-110-083605}.
On the other hand, compared with the conventional OMIT \cite{pra-83-043826},
our double OMIT characterizes as a linear variation of the peak value with
respect to the environmental temperature, which is robust to the cavity
decay and does not vary with the Coulomb coupling between the two
nano-mechanical resonators (NAMRs). This feature exists no matter whether the
two NAMRs are identical or not, which is useful for practical applications as discussed
later.

The paper is organized as follows. In Sec. II, we present the solution to
the model of our designed double OMIT and focus on the spectra via a
homodyne detection. In Sec. III, some calculations are made numerically with
experimentally available values, justifying some unique features, such as
robustness against the cavity decay and invariance with the Coulomb coupling
strength. The feasibility of precision measurement of the environmental
temperature is discussed in Sec. IV. Other extended discussion is made in
Sec.V and a brief conclusion is given in the last section.

\section{The model and solution}

\begin{figure}[tph]
\center
\includegraphics[width=1\columnwidth]{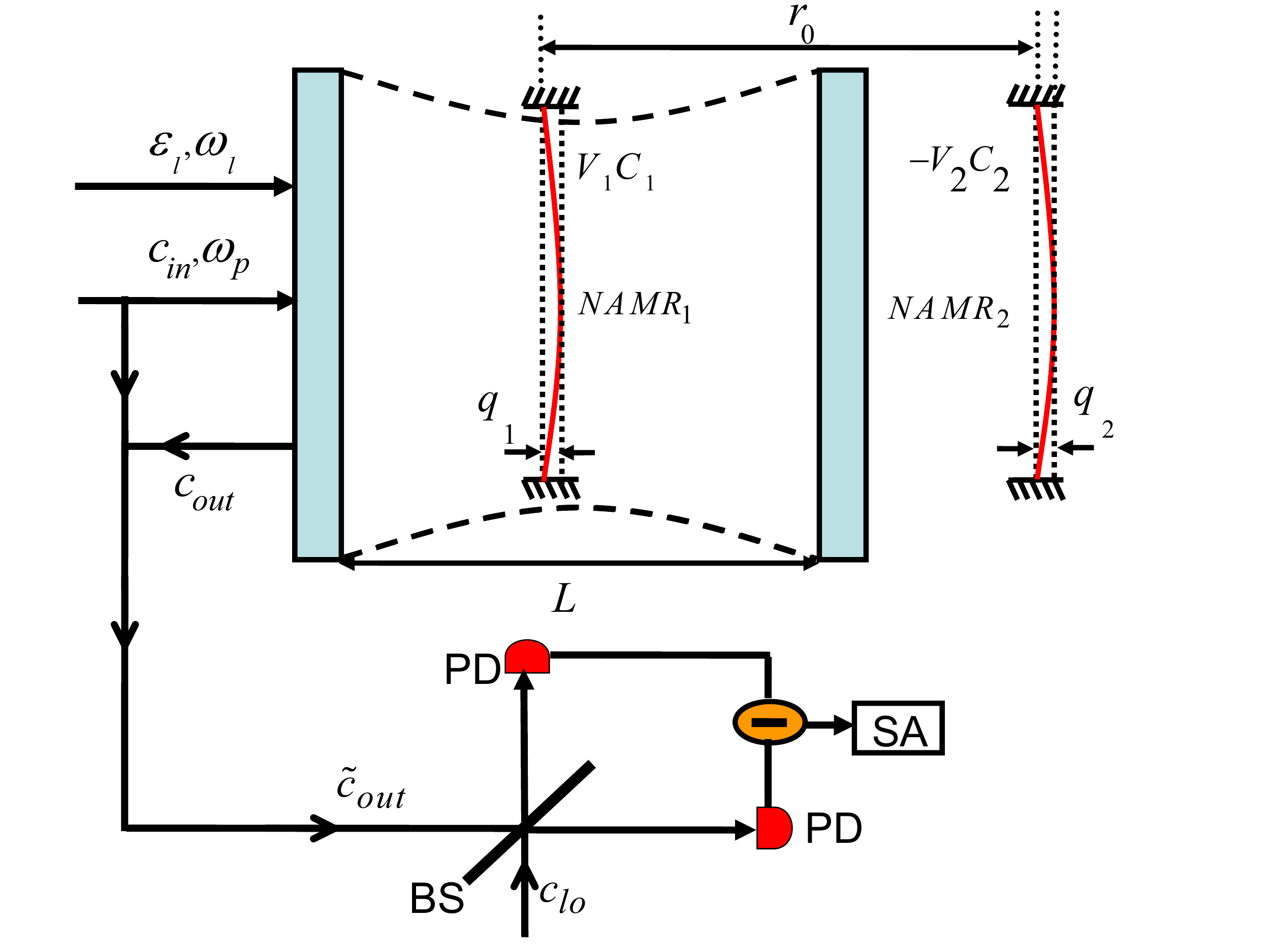}
\caption{(Color online) Schematic diagram of the double OMIT system and the
measurement. A high-quality Fabry-Per$\acute{o}$t cavity consists of two
fixed mirrors and a charged NAMR$_{1}$, which is charged by the bias gate
voltage $V_{1}$ and subject to the Coulomb force due to the charged NAMR$%
_{2} $ outside the cavity and with the bias gate voltage $V_{2}$. The
optomechanical cavity of length $L$ is driven by two light fields, one of
which is the pump field $\protect\varepsilon_{l}$ with frequency $\protect%
\omega _{l}$ and the other of which is the probe field $c_{in}$ with
frequency $\protect\omega_{p}$. $q_{1}$ and $q_{2}$ represent, respectively,
the small displacements of the two NAMRs from their equilibrium positions,
with $r_{0}$ the equilibrium distance between them. The output field $%
c_{out} $ from the cavity turns to be $\tilde{c}_{out}$, which is mixed with
a strong local field $c_{lo}$ centered around the probe frequency $\protect%
\omega _{p}$ at a 50:50 beam splitter (BS). Finally the homodyne spectra are
obtained by the spectrum analyzer (SA) assisted by two photon detectors
(PD). }
\end{figure}
As sketched in Fig. 1, there are two charged NAMRs with one (NAMR$_{1}$)
inside a Fabry-Per$\acute{o}$t (FP) cavity coupling to the cavity mode by
the radiation pressure and interacting with the other (NAMR$_{2}$) outside
the cavity. The FP cavity contains two mirrors distant by $L$ with the
left-hand side mirror partially transmitting and the right-hand side one $%
100\%$ reflecting. There is a driving on the cavity mode from the left-hand
side mirror by a strong coupling field with frequency $\omega_{l}$. The
system can be described as
\begin{eqnarray}
H_{whole} &=&\hbar \omega _{c}c^{\dagger }c+\sum_{j=1}^{2}(\frac{p_{j}^{2}}{
2m_{j}}+\frac{1}{2}m_{j}\omega _{j}^{2}q_{j}^{2})  \label{1} \\
&-&\hbar gc^{\dagger }cq_{1}+H_{I}+i\hbar \varepsilon _{l}(c^{\dagger
}e^{-i\omega _{l}t}-H.C.),  \nonumber
\end{eqnarray}
where the first term is for the single-mode cavity field with frequency $%
\omega_{c}$ and annihilation (creation) operator $c (c^{\dagger})$. The
second and third terms describe the vibration of the charged NAMRs with
frequency $\omega_{1} (\omega_{2})$ and effective mass $m_{1} (m_{2})$. $%
p_{1} (p_{2})$ and $q_{1} (q_{2})$ are the momentum and the position
operators of NAMR$_{1}$ (NAMR$_{2}$), respectively. The fourth term presents
the radiation pressure coupling the cavity field to the NAMR$_{1}$ with a
coupling strength $g=\omega_{c}/L$.

Coulomb coupling between the two charged NAMRs is given by $H_{I}=\frac{%
-C_{1}V_{1}C_{2}V_{2}}{4\pi \varepsilon _{0}|r_{0}+q_{1}-q_{2}|}$, where $%
r_{0}$ is the distance between the equilibrium positions, and NAMR$_{1}$ and
NAMR$_{2}$ take the charges $C_{1}V_{1}$ and $-C_{2}V_{2}$ with $C_{1}$ and $%
C_{2}$ being the capacitance of the gates, respectively. Under the
assumption that the deformations of the NAMRs are much less than their
distance ($q_{1},q_{2}\ll r_{0}$), the Hamiltonian $H_{I}$ can be expanded
to the second order as $H_{I}=\frac{-C_{1}V_{1}C_{2}V_{2}}{4\pi \varepsilon
_{0}r_{0}}[1-\frac{q_{1}-q_{2}}{r_{0}}+(\frac{q_{1}-q_{2}}{r_{0}})^{2}]$.
Since the linear term may be absorbed into the definition of the equilibrium
positions, and the quadratic term includes the renormalized oscillation
frequencies for both the NAMRs, we have a reduced form $H_{I}=\hbar \lambda
q_{1}q_{2}$ for $\lambda = \frac{C_{1}V_{1}C_{2}V_{2}}{2\pi \hbar
\varepsilon _{0}r_{0}^{3}}$ \cite{pra-72-041405,prl-93-266403}.

The last term represents the cavity field driven by an input field with
frequency $\omega_{l}$, where the pump field strength $\varepsilon _{l}=%
\sqrt{2\kappa\wp/\omega_{l}}$ depends on the power $\wp$ of the coupling
field and the cavity decay rate $\kappa$.

In a frame rotating with the pump field frequency $\omega_{l}$, the
Hamiltonian in Eq. (\ref{1}) is rewritten as
\begin{eqnarray}
H_{total} &=&\hbar \Delta _{c}c^{\dagger }c+\sum_{j=1}^{2}(\frac{p_{j}^{2}}{
2m_{j}}+\frac{1}{2}m_{j}\omega _{j}^{2}q_{j}^{2})  \nonumber \\
&-&\hbar gc^{\dagger }cq_{1}+\hbar \lambda q_{1}q_{2}+i\hbar \varepsilon
_{l}(c^{\dagger }-c),  \label{2}
\end{eqnarray}
with $\Delta _{c}=\omega _{c}-\omega _{l}$. Considering the decay rates $%
\gamma _{1}$ and $\gamma_{2}$ for the NAMR$_{1}$ and NAMR$_{2}$,
respectively, we obtain the corresponding frequency-domain correlation
functions for the thermal noise $\xi _{1}$ and $\xi _{2}$ at a temperature $%
T $,
\[
\langle \xi _{\tau }(\omega )\xi _{\tau }(\Omega )\rangle =2\pi \hbar \gamma
_{\tau }m_{\tau }\omega \left[ 1+\coth \left( \frac{\hbar \omega }{2k_{B}T}%
\right)\right] \delta (\omega +\Omega),
\]
where $\tau =1,2$ and $k_{B}$ is the Boltzmann constant.

We assume that the cavity mode $c$ couples to the input quantum field $%
c_{in} $, which is a narrow-band squeezed field with the center around the
frequency $\omega _{p}=\omega _{c}+\omega _{1}$ and with a finite bandwidth $%
\Gamma$. The nonvanishing correlation functions for this input squeezed
field are given by
\begin{eqnarray}
\langle c_{in}(\omega )c_{in}(\Omega )\rangle &=&2\pi \frac{M\Gamma ^{2}}{%
\Gamma ^{2}+(\omega -\omega _{1})^{2}}\delta (\omega +\Omega -2\omega _{1}),
\nonumber \\
\langle c_{in}(\omega )c_{in}^{\dagger }(-\Omega )\rangle &=&2\pi \left[
\frac{N\Gamma ^{2}}{\Gamma ^{2}+(\omega -\omega _{1})^{2}}+1\right] \delta
(\omega +\Omega ),  \nonumber \\
\end{eqnarray}
where $N$ is the photon number in the squeezed vacuum, and $M=\sqrt{N(N+1)}$
is an anti-normally ordered term including a broadband contribution from the
vacuum noise.

Considering the input squeezed field, Eq. (\ref{2}) under dissipation and
fluctuation is governed by quantum Langevin equations, yielding
\begin{eqnarray}
\dot{q}_{1} &=&\frac{p_{1}}{m_{1}},~~~~~~~~\dot{q}_{2}=\frac{p_{2}}{m_{2}},
\nonumber \\
\dot{c} &=&-[\kappa +i(\Delta _{c}-gq_{1})]c+\varepsilon _{l}+\sqrt{2\kappa}
c_{in},  \nonumber \\
\dot{p}_{1} &=&-m_{1}\omega _{1}^{2}q_{1}-\hbar \lambda q_{2}+\hbar
gc^{\dagger }c-\gamma _{1}p_{1}+\xi _{1},  \nonumber \\
\dot{p}_{2} &=&-m_{2}\omega _{2}^{2}q_{2}-\hbar \lambda q_{1}-\gamma
_{2}p_{2}+\xi_{2},
\end{eqnarray}
whose steady solutions are given by,
\begin{eqnarray}
p_{1s} &=&p_{2s}=0,~~~~q_{1s}=\frac{\hbar g|c_{s}|^{2}}{m_{1}\omega
_{1}^{2}- \frac{\hbar ^{2}\lambda ^{2}}{m_{2}\omega _{2}^{2}}},  \nonumber \\
q_{2s} &=&\frac{\hbar \lambda q_{1s}}{-m_{2}\omega _{2}^{2}},~~~~c_{s}=\frac{
\varepsilon _{l}}{\kappa +i\Delta},
\end{eqnarray}
with $\Delta =\Delta _{c}-gq_{1s}$ being the effective detuning between the
cavity and the driven fields.

Next,we consider the linear operators as steady mean values plus additional
fluctuation operators,
\[
q_{\tau }=q_{\tau s}+\delta q_{\tau },~~p_{\tau }=p_{\tau s}+\delta p_{\tau
},~~c=c_{s}+\delta c,
\]
where $\delta q_{\tau }$, $\delta p_{\tau }$, $\delta c$ are small
fluctuations around the corresponding steady values. Since the steady values
have no contribute on the output fields, we only focus on the fluctuation
operators, which work as the probe fields and influence the output fields.

Defining the fluctuation operators $\mathbf{X}=(\delta p_{1}(\omega ),\delta
q_{1}(\omega ),\delta p_{2}(\omega ),\delta q_{2}(\omega ),\delta
c(\omega),\delta c^{\dagger }(\omega ))^{T}$, we reach the linearized
quantum Langevin equations $\mathbf{A}\mathbf{X}=\mathbf{B}$ for the
fluctuation operators from Eq. (4), where
\begin{widetext}
\begin{eqnarray}
\textbf{A}=\left(
\begin{array}{cccccc}
1 & i\omega m_{1} & 0 & 0 & 0 & 0  \\
-i\omega+\gamma_{1} & m_{1}\omega_{1}^{2} & 0 & \hbar\lambda & -\hbar gc_{s}^{*}  & -\hbar gc_{s} \\
0 & 0 & 1 & i\omega m_{2} & 0 & 0 \\
0 & \hbar\lambda & -i\omega+\gamma_{2} & m_{2}\omega_{2}^{2} & 0 & 0 \\
0 & -igc_{s} & 0 & 0 & \kappa+i(\Delta-\omega) & 0  \\
0 & igc_{s}^{*} & 0 & 0 & 0 & \kappa-i(\Delta+\omega)  \\
\end{array}
\right),
\end{eqnarray}
\end{widetext}and $\mathbf{B}=(0,\xi _{1}(\omega ),0,\xi _{2}(\omega ),\sqrt{
2\kappa}c_{in}(\omega ),\sqrt{2\kappa}c_{in}^{\dagger}(-\omega))^{T}$. Thus
the fluctuation $\delta c(\omega )$ of the cavity field can be solved by the
linearized equations above.

Based on the input-output relation $c_{out}(\omega )=\sqrt{2\kappa }c(\omega
)-c_{in}(\omega )$, we define the output field as $\tilde{c}_{out}(\omega
)=c_{out}(\omega )+c_{in}(\omega )=\sqrt{2\kappa }c(\omega )$ in order to
study the physical properties of the light field leaking out of the cavity,
as in \cite{pra-86-013815,pra-83-043826}. Straightforward deduction yields
\begin{eqnarray}
\delta \tilde{c}_{out}(\omega ) &=&E(\omega )c_{in}(\omega )+F(\omega
)c_{in}^{\dagger }(-\omega )  \nonumber \\
&+&V_{1}(\omega )\xi _{1}(\omega )+V_{2}(\omega )\xi _{2}(\omega ),
\end{eqnarray}%
where
\begin{eqnarray}
E(\omega ) &=&2\kappa \lbrack \frac{1}{\kappa +i(\Delta -\omega )}  \nonumber
\\
&&+\frac{i\hbar g^{2}|c_{s}|^{2}(\Delta +\omega +i\kappa )m_{2}B_{1}}{%
(\Delta -\omega -i\kappa )d(\omega )}], \\
F(\omega ) &=&\frac{-2i\kappa \hbar g^{2}c_{s}^{2}m_{2}B_{1}}{d(\omega )},
\nonumber \\
V_{1}(\omega ) &=&\frac{\sqrt{2\kappa }ic_{s}g[-\kappa +i(\Delta +\omega
)]m_{2}B_{1}}{d(\omega )},  \nonumber \\
V_{2}(\omega ) &=&\frac{-\sqrt{2\kappa }\hbar gc_{s}\lambda (\Delta +\omega
+i\kappa )}{d(\omega )},
\end{eqnarray}%
with $d(\omega )=-\hbar ^{2}\lambda ^{2}A+m_{2}B_{1}(2|c_{s}|^{2}g^{2}\hbar
\Delta +m_{1}AB_{2})$, $A=\Delta ^{2}+(\kappa -i\omega )^{2}$, $B_{1}=\omega
^{2}+i\omega \gamma _{2}-\omega _{2}^{2}$ and $B_{2}=\omega ^{2}+i\omega
\gamma _{1}-\omega _{1}^{2}$.

Based on the above mentioned correlation functions of $c_{in}(\omega)$ and $%
\xi_{\tau}(\omega)$ as well as the standard homodyne detection \cite%
{Introductory Quantum Optics} as plotted in Fig. 1, we may understand
characteristics of the system from the homodyne spectrum $X(\omega)$, which
can be analytically expressed as below provided that the fast oscillating
terms at frequencies $\pm 2\omega _{1(2)}$ are omitted,
\begin{widetext}
\begin{eqnarray}
X(\omega)&=&E(\omega+\omega_{1})E(-\omega+\omega_{1})\frac{M\Gamma^{2}}{\Gamma^{2}+\omega^{2}}+|E(\omega+\omega_{1})|^{2}
\frac{N\Gamma^{2}}{\Gamma^{2}+\omega^{2}} \nonumber\\
&+&E^{*}(-\omega+\omega_{1})E^{*}(\omega+\omega_{1})\frac{M\Gamma^{2}}{\Gamma^{2}+\omega^{2}}
+|E(-\omega+\omega_{1})|^{2}\frac{N\Gamma^{2}}{\Gamma^{2}+\omega^{2}} \nonumber\\
&+&|E(\omega+\omega_{1})|^{2}+|F(-\omega+\omega_{1})|^{2} \nonumber\\
&+&|V_{1}(\omega+\omega_{1})|^{2}\hbar\gamma_{1}m_{1}(\omega+\omega_{1})
\times\left\{1+\coth\left[\frac{\hbar(\omega+\omega_{1})}{2k_{B}T}\right]\right\}\nonumber\\
&+&|V_{2}(\omega+\omega_{1})|^{2}\hbar\gamma_{2}m_{2}(\omega+\omega_{1})
\times\left\{1+\coth\left[\frac{\hbar(\omega+\omega_{1})}{2k_{B}T}\right]\right\}\nonumber\\
&+&|V_{1}(-\omega+\omega_{1})|^{2}\hbar\gamma_{1}m_{1}(\omega-\omega_{1})
\times\left\{1+\coth\left[\frac{\hbar(\omega-\omega_{1})}{2k_{B}T}\right]\right\}\nonumber\\
&+&|V_{2}(-\omega+\omega_{1})|^{2}\hbar\gamma_{2}m_{2}(\omega-\omega_{1})
\times\left\{1+\coth\left[\frac{\hbar(\omega-\omega_{1})}{2k_{B}T}\right]\right\}, \label{eq014}
\end{eqnarray}
\end{widetext}
where the first four terms are from the input squeezed field and the next
two terms are relevant to the spontaneous emission of the input vacuum
noise. The rest terms are caused by the thermal noise of the NAMRs, which is
temperature dependent. So the squeezed field employed does not work better for
the temperature-dependent effects in comparison to other quantum fields, but enhancing the measurement precision
in the homodyne spectrum. In addition, Eq. (\ref{eq014}) is more general
with respect to the counterpart in Ref. \cite{pra-83-043826}, since it can
be reduced to Eq. (13) in \cite{pra-83-043826} if $\lambda =0$, i.e., in the
absence of the Coulomb coupling.

\section{homodyne spectra of the Double OMIT}

We specify below some unique characteristics of the double OMIT by the
numerically calculated homodyne spectra. For simplicity, we first suppose
the two NAMRs to be identical in our treatment. The non-identical case,
which is more general but not fundamentally different, will be justified
later.

Our numerical calculation is carried out based on realistic parameter values
\cite{Nature-460-724}. We consider an optomechanical cavity with length $%
L=25 $ mm and decay rate $\kappa\sim 2\pi\times 215$ kHz, driven by the pump
field of wavelength $\lambda _{l}=2\pi c/\omega_{l}=1064$ nm. For the two
identical NAMRs, we assume the effective mass $m_{1}=m_{2}=145$ ng, the
eigen-frequencies $\omega_{m}=\omega_{1}=\omega _{2}=2\pi\times 947$ kHz,
the decay rates $\gamma_{m}=\gamma_{1}=\gamma _{2}=2\pi\times 141$ Hz, and
the quality factors $Q_{1}=Q_{2}=\omega_{m}/\gamma_{m}=6700$. In addition,
the linewidth of the squeezed vacuum is supposed to be $\Gamma =2\kappa$.

\begin{figure}[tph]
\centering
\includegraphics[width=1 \columnwidth]{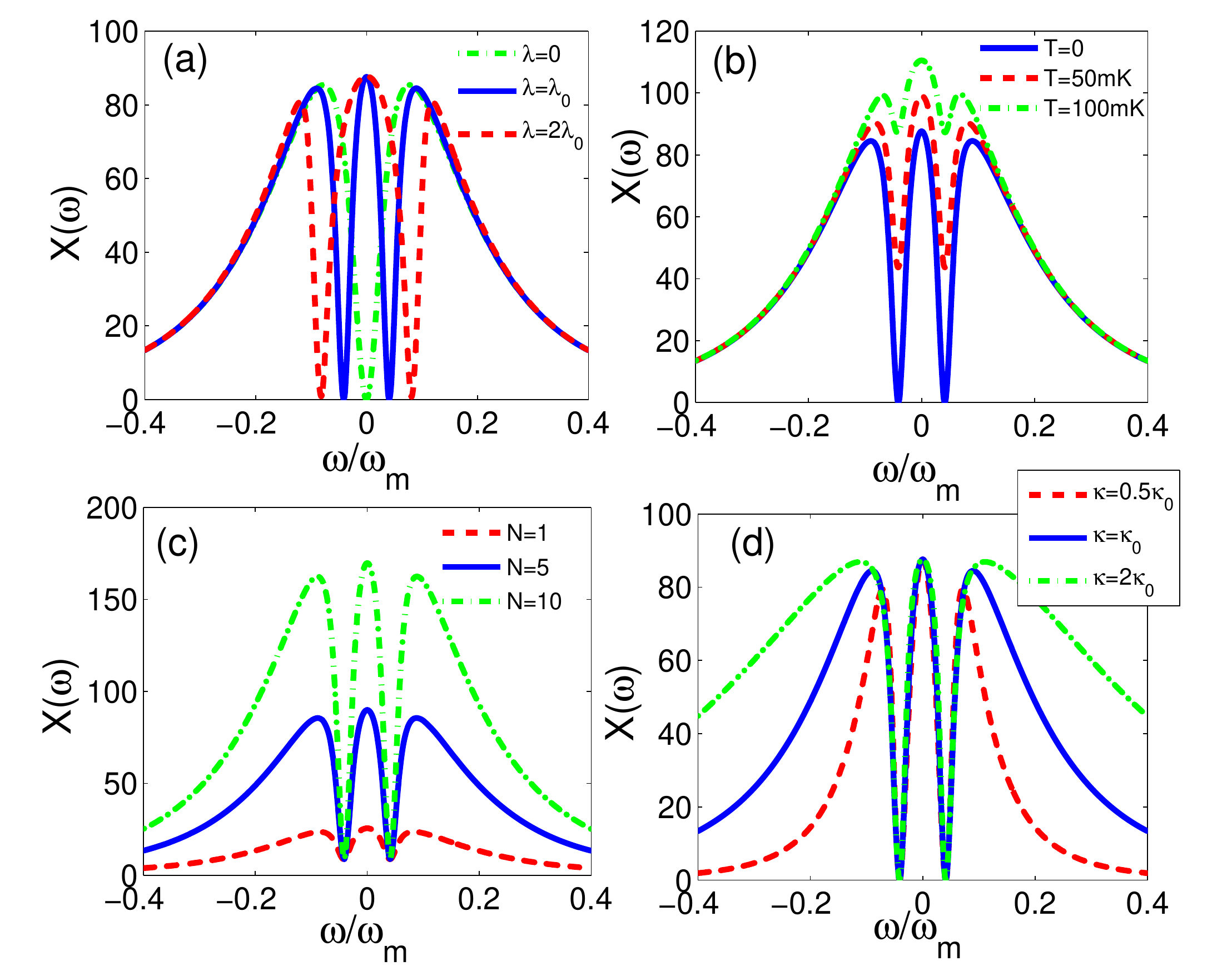}
\caption{(Color online) The homodyne spectra $X(\protect\omega)$ of the
output field as functions of the normalized frequency $\protect\omega /%
\protect\omega_{m}$, where (a) for different Coulomb coupling $\protect%
\lambda$ with the coupling strength unit $\protect\lambda_{0}=4\times
10^{36} $ Hz/m$^{2}$; (b) for different temperature T with N=5; (c) for
different photon number $N$ with $T=10$ mK; (d) for different cavity decay $%
\protect\kappa$ with the decay unit $\protect\kappa_{0}=2\protect\pi%
\times215 $ kHz.}
\label{fig2}
\end{figure}

In most of the calculations below, we employ the zero temperature $T=0$ and
the photon number $N=5$ in the squeezed vacuum, and assume the coupling
field power $\wp =2$ mW and the coupling strength unit $\lambda_{0}=4\times
10^{36}$ Hz/m$^{2}$. The homodyne spectrum $X(\omega)$ plotted in Fig. \ref%
{fig2}(a) presents the change from a single transparency window to two
transparency windows with increasing Coulomb coupling, which reflects a fact
that the Coulomb coupling breaks down the original interference in the OMIT
and splits the bosonic mode of the system into two. Since the energy
difference between the two split bosonic modes depends on the Coulomb
coupling, the splitting of the transparency windows is relevant to the
Coulomb coupling \cite{pra-90-023817}. However, the middle peak is fixed no
matter how much the Coulomb coupling varies.

For a finite temperature, the homodyne spectrum $X(\omega)$ still works for
the double OMIT, as presented in Fig. 2(b) where the visible middle peak and
two nadirs exist even at $T=100$ mK. However, the trend reflected in Fig.
2(b) indicates that the double-OMIT will definitely disappear with further
increase of the temperature. Besides, a more careful calculation shows that
the photon number plays an important role in the $X(\omega)$ variation. From
Fig. 2(c), we find that the middle peak and two nadirs are visible at $T=10$
mK, even for the squeezed state at the single-photon level (the red dashed
curve). In particular, compared with Fig. 2(b), we find that the two nadirs
of the double OMIT are fixed with the variation of the photon number, but
changing with the temperature.

Moreover, the cavity decay modifies the profiles of the transparency
windows, as plotted in Fig. \ref{fig2}(d). However, although the profiles
become narrower and sharper with smaller cavity decay rate $\kappa$, the
peak and the nadirs of the two transparency windows remain unchanged,
implying robustness against the cavity decay at these points.

\begin{figure}[tph]
\center
\includegraphics[width=1 \columnwidth]{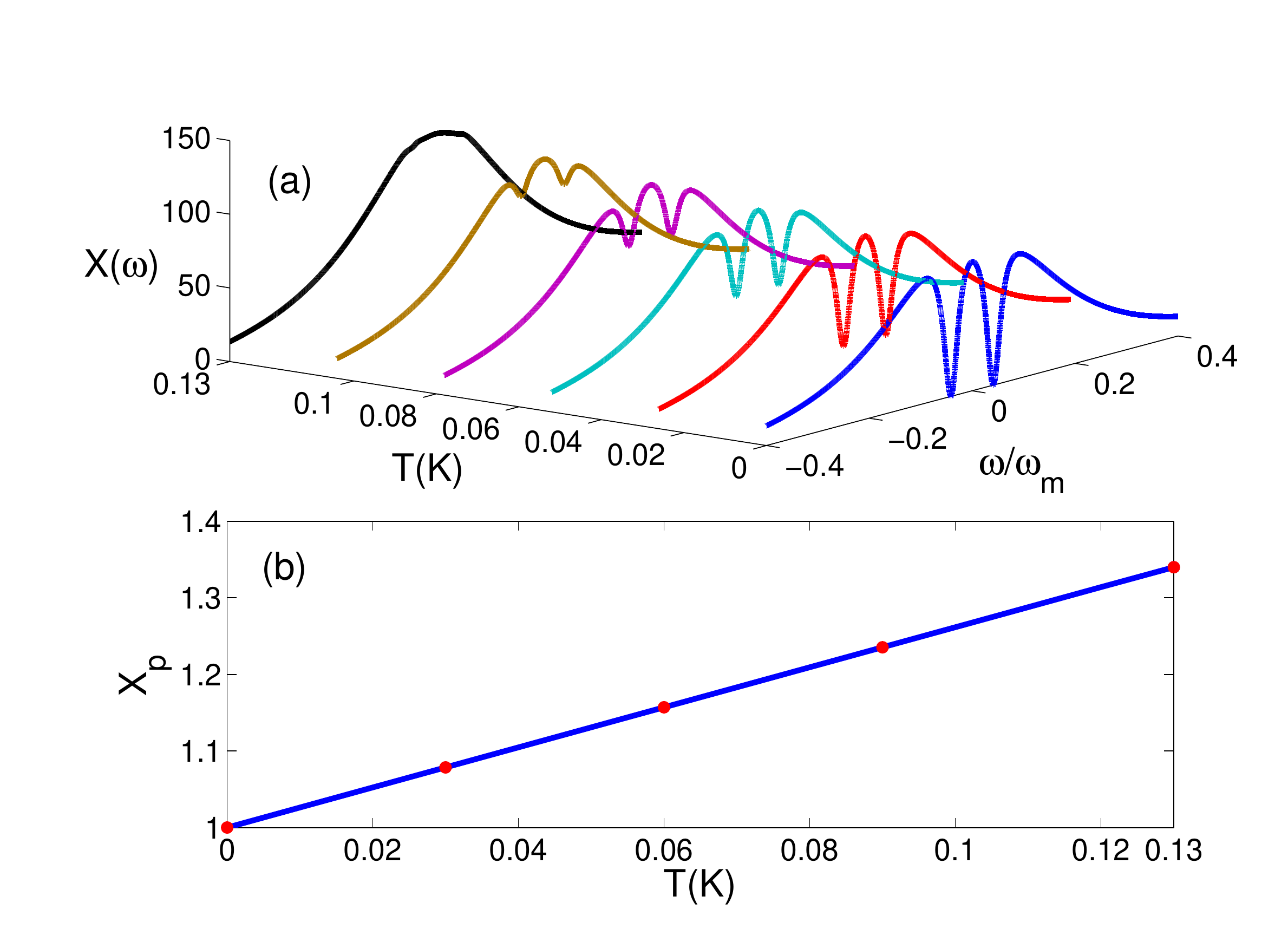}
\caption{(Color online) (a) The homodyne spectra $X(\protect\omega)$ as
functions of the frequency $\protect\omega /\protect\omega_{m}$ and the
temperature $T$ for $\protect\lambda=\protect\lambda_{0}$. (b) The rescaled
middle peak value $X_{p}$ (in units of $X_{0}=87.59$) as a function of the
temperature $T$ for $\protect\lambda=\protect\lambda_{0}$, where the
temperature measurement is available within the range $0\le T\le$0.13 K. The
upper limit of the measured temperature is restricted by the measurement
resolution $1\%$ of $X_{p}$ at $T=$0.13 K. Other parameters take the same
values as in Fig. 2(a).}
\label{fig3}
\end{figure}

\begin{figure}[tph]
\center
\includegraphics[width=1 \columnwidth]{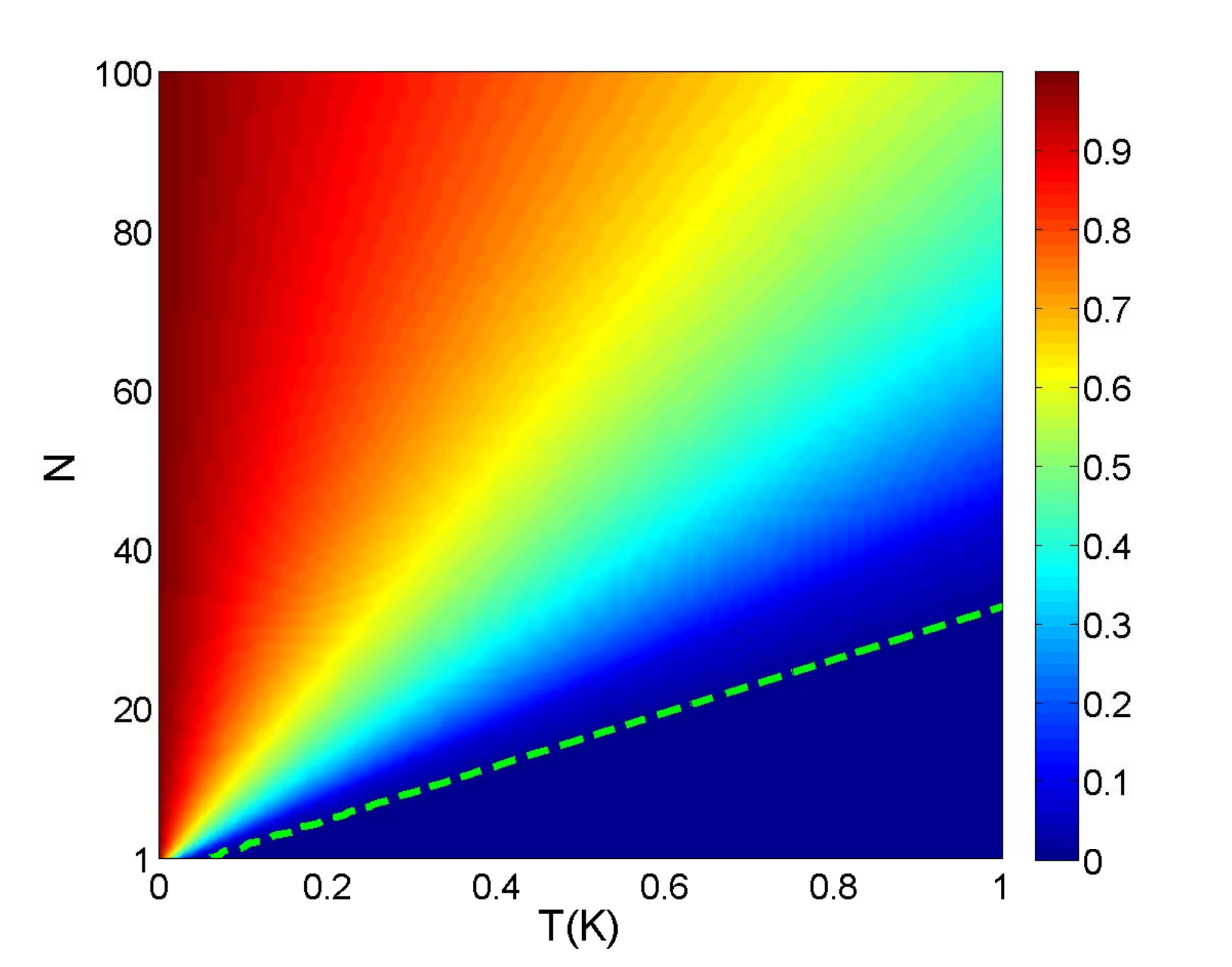}
\caption{(Color online) The quantum signal visibility $(QSV)$ as functions
of the photon number $N$ and temperature $T$ for $\protect\lambda=\protect%
\lambda_{0}$. The dashed line in green represents a borderline for available
measurement, below which $QSV$ is less than the measurement resolution $1\%$%
. Other parameters are of the same values as in Fig. 2(a).}
\label{fig4}
\end{figure}

\section{MEASUREMENT OF THE environmental temperature}

Using above indicated properties of the homodyne spectrum $X(\omega)$, we
may carry out precision measurement of the environmental temperature using
the double-OMIT with the squeeze field.

Fig. \ref{fig3}(a) presents the step-by-step change of $X(\omega)$ with the
environmental temperature, where the middle peak increases linearly with the
temperature. Since we have renormalized the middle peak value by $%
X_p=X(0)/X_0$ with $X_0=87.59=X(0)$ at zero temperature, the temperature
change can be exactly known from Fig. 3(b) by precisely measuring the
variation of $X_{p}$. But as shown in Fig. 3(a), the height difference
between the middle peak and the two nadirs shrinks with the temperature
increasing. This implies an upper limit of the measured temperature, e.g., $
T=0.13$ K where the double OMIT reaches the resolution limit of the
observation.

Specifically, for the linear variation of $X_p$ with respect to $T$ in Fig.
\ref{fig3}(b), the sensitivity can be evaluated by the slope $k=\partial
T/\partial X(0)=4.4\times 10^{-3}$K. As a result, for a measurement
resolution $\delta {X(0)}=1\%$ of the peak value, the detectable temperature change
can be minimized to  $4.4\times10^{-5}$ K, which is lower by one
order of magnitude compared to a quantum thermometer designed based on a
noise measurement using the electron charge \cite{science-300-1929}. The
linear variation of the middle peak with respect to the temperature can be
understood from Eq. (10) in which the last four terms are reduced to be
linearly changing with $T$ if $T\rightarrow$0. In fact, the measurement precision in
our case can be further enhanced if we elaborately change the mass ratio of
the two NAMRs, as discussed later. We have to emphasize that this
measurement based on the peak values is insensitive to the change of Coulomb
coupling, as indicated in Fig. 2(a).

To carry out a precision measurement in our scheme, we have to have a big
enough contrast of $X(\omega)$ for our observation. To this end, we
consider below the influence from the photon number and the environmental
temperature. When the two NAMRs are identical, if many photons are involved in
the squeezed state, we have $N\approx M$, and Eq. (10) at low temperature ($
T\rightarrow 0$) expands to the first order at the frequency of the
peak point ($\omega _{peak}\equiv 0$) as
\begin{widetext}
\begin{eqnarray}
X(\omega _{peak})=X(0)&=&N[E(\omega _{m})+E^{\ast }(\omega_{m})]^{2}+|E(\omega _{m})|^{2}+|F(\omega _{m})|^{2} \nonumber \\
&+&2|V_{1}(\omega _{m})|^{2}\hbar \gamma
_{1}m_{1}\omega _{m}\coth [\frac{\hbar \omega _{m}}{2k_{B}T}]  \nonumber \\
&+&2|V_{2}(\omega _{m})|^{2}\hbar \gamma _{2}m_{2}\omega _{m}\coth [\frac{%
\hbar \omega _{m}}{2k_{B}T}]  \nonumber \\
&=&N[E(\omega _{m})+E^{\ast }(\omega _{m})]^{2}+|E(\omega
_{m})|^{2}+|F(\omega _{m})|^{2}+ k(0)T,
\end{eqnarray}
\end{widetext}
where the slope is $k(0)=4k_{B}\gamma_{m}m[|V_{1}(\omega_{m})|^{2}+|V_{2}(\omega_{m})|^{2}$], $N$ is relevant to
the quantum signal terms which compete with the last two thermal noise terms
involving $T$. With the increase of $T$, the values of the thermal noise
terms will exceed those of the quantum signal terms and thus the quantum
signal is completely buried by the thermal noise, i.e., disappearance of the
double OMIT. To clarify this point, we define quantum signal visibility $%
QSV=(\mathtt{peak}-\mathtt{nadir})/\mathtt{peak}$ as a contrast of our
observation, where $QSV=1$ implies an overwhelming quantum signal and
disappearance of the double OMIT corresponds to $QSV=0$. Involvement of more
photons helps increasing the contrast, as indicated in Fig. 2(c) and
understood from Eq. (11). We plot in Fig. 4 a borderline for available
precision measurement, below which $QSV$ is smaller than $1\%$, the
measurement resolution we assumed above. As such, if $T=0$, the double OMIT
always exists no matter how many photons are involved. But with $T$
increasing, more photons are required to be involved if our proposed
measurement of the environmental temperature works.

\section{Discussion}

\begin{figure}[tph]
\center
\includegraphics[width=1 \columnwidth]{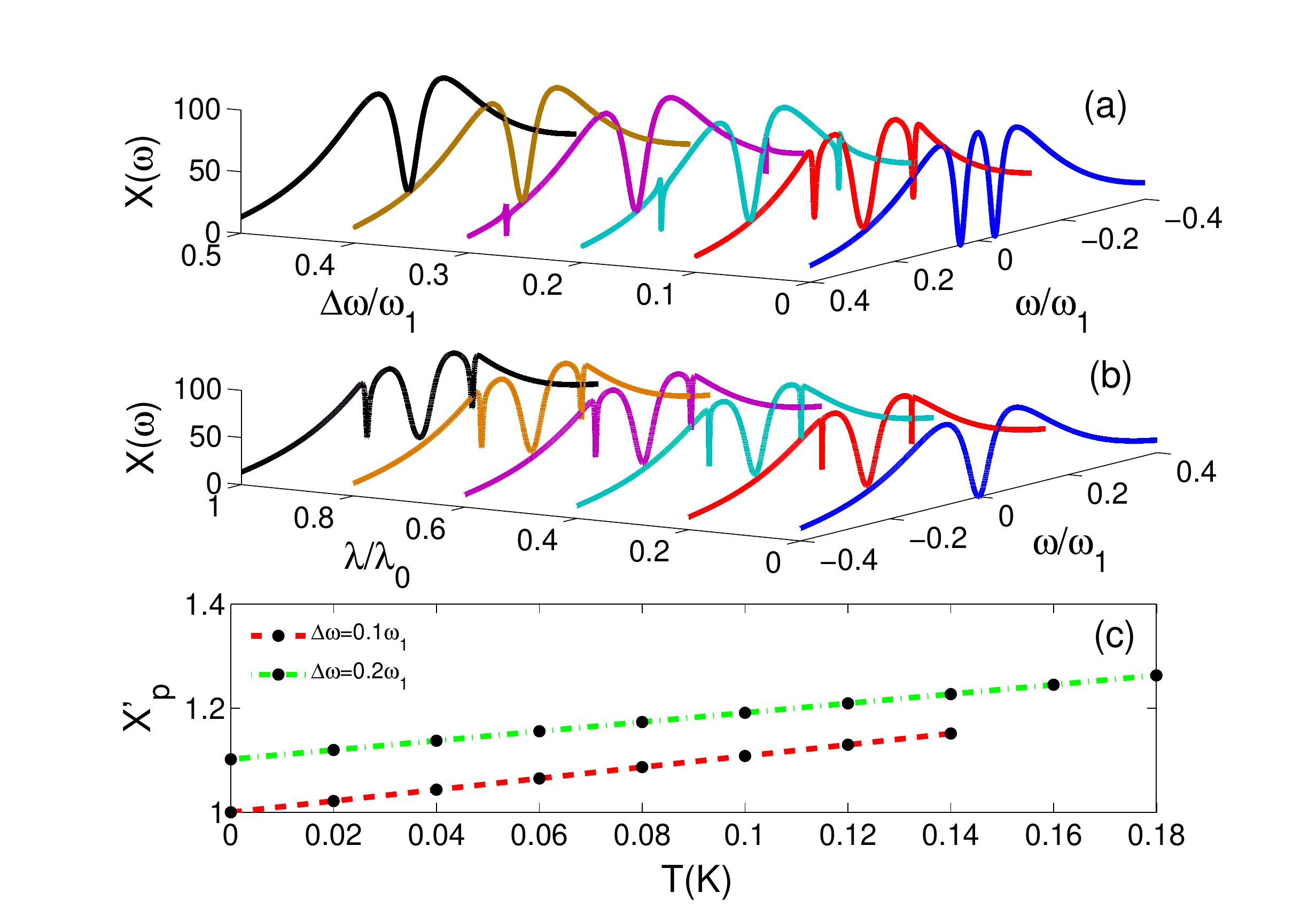}
\caption{(Color online) (a) The homodyne spectra $X(\protect\omega)$ as
functions of $\protect\omega/\protect\omega_{1}$ and $\Delta\protect\omega/%
\protect\omega_{1}$ for $\protect\lambda=\protect\lambda_{0}$. (b) The
homodyne spectra $X(\protect\omega)$ as functions of $\protect\omega/\protect%
\omega_{1}$ and $\protect\lambda/\protect\lambda_{0}$ with $\Delta\protect%
\omega=0.1\protect\omega_{1}$. (c) The rescaled peak value $X^{\prime}_{p}$
(in units of 84.03) as a function of the temperature T for $\protect\lambda=%
\protect\lambda_{0}$, where the available measurement of the temperature is $%
0\le T \le$0.14 K ($\Delta\protect\omega = 0.1\protect\omega_1$) or $0\le
T\le$0.18 K ($\Delta\protect\omega = 0.2\protect\omega_1$). The upper limit
of the measured temperature is restricted by the measurement resolution $1\%$
of $X^{\prime}_{p}$. Other parameters are of the same values as in Fig.
2(a). }
\label{fig5}
\end{figure}

\begin{figure}[tph]
\center
\includegraphics[width=1 \columnwidth]{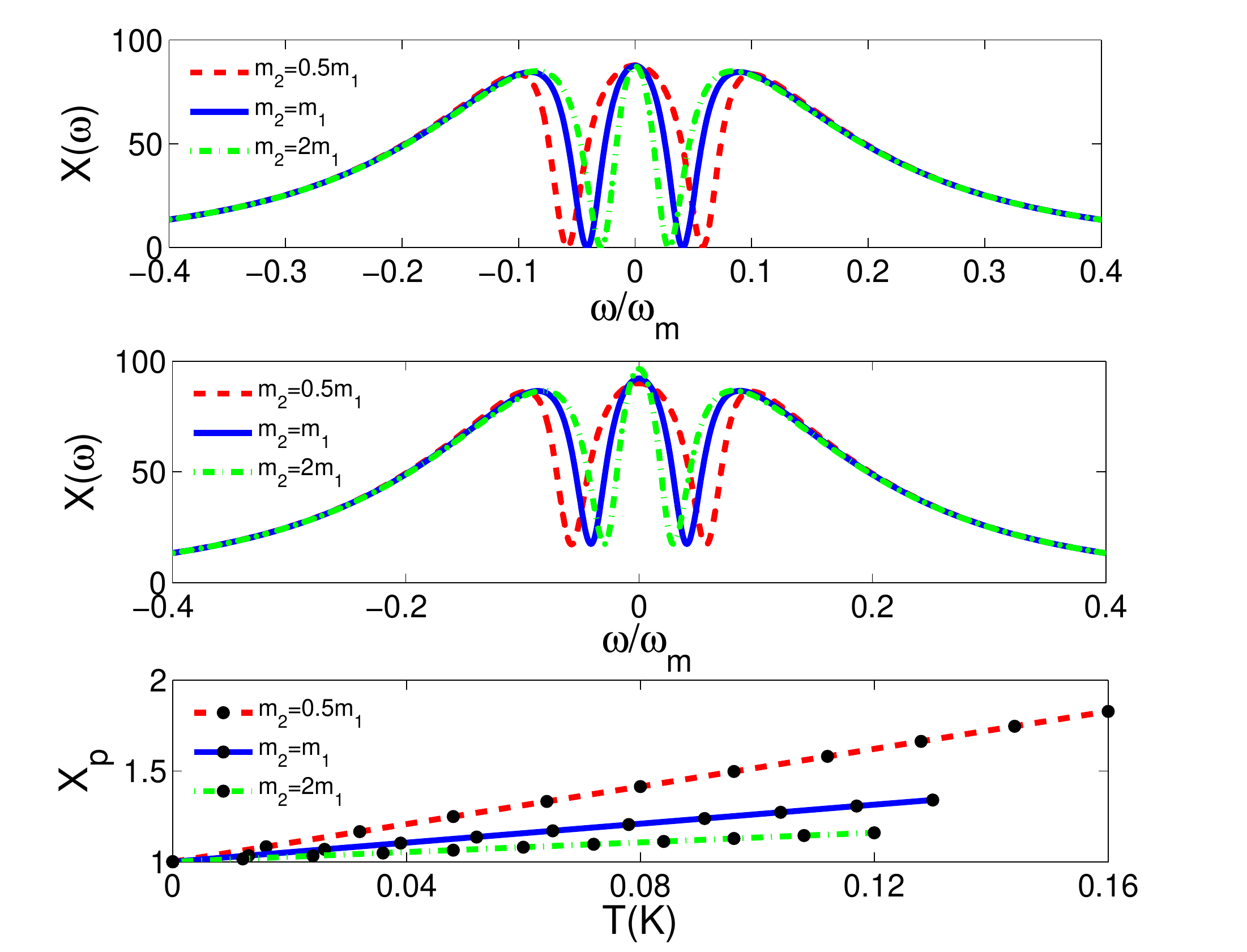}
\caption{(Color online) (a) The homodyne spectra $X(\protect\omega)$ as
functions of $\protect\omega/\protect\omega_{m}$ for $\protect\lambda=%
\protect\lambda_{0}$ and T=0. (b) The homodyne spectra $X(\protect\omega)$
as functions of $\protect\omega/\protect\omega_{m}$ for $\protect\lambda=%
\protect\lambda_{0}$ and $T=20$ mK. (c) The rescaled middle peak value $%
X_{p} $ (in units of 87.59) as a function of the temperature T for $\protect%
\lambda=\protect\lambda_{0}$, where the temperature measurement is available
within the range $0\le T\le$0.16 K ($m_2 = 0.5 m_1$) or $0\le T \le$0.12 K ($%
m_2 = 2m_1$). The upper limit of the measured temperature is restricted by
the measurement resolution $1\%$ of $X_{p}$. Other parameters take the same
values as in Fig. 2(a).}
\label{fig5}
\end{figure}
The precision measurement of the environmental temperature described above
is based on identical NAMRs. A more general and realistic case is with
non-identical NAMRs, which are different in frequency or mass. In such
cases, our model has different characteristics and thus different
applications.

As an example, we first consider in Fig. 5 the situation with different
frequencies of the two NAMRs. In this case, the profile of the double OMIT
keeps changing with the frequency difference $\Delta \omega =|\omega
_{1}-\omega _{2}|$, where the double windows are first split into triple
windows and then become a standard OMIT with a single window, like the
absence of Coulomb coupling (See Fig. 5(a)). Fig. 5(b) provides another
view angle to observe the role played by the Coulomb coupling. Different
from the situation with identical NAMRs, the Coulomb coupling yields triple
windows directly from the single window once the Coulomb coupling
turns to be non-zero. This feature is actually resulted by the interference
of two double OMITs with two asymmetric windows due to the frequency
difference. As indicated by the homodyne spectra in Figs. 5(a) and 5(b),
although the profiles of the spectra change in the variation of $\Delta
\omega $ and $\lambda $, the two symmetric peaks in the case of triple
windows are always fixed, where the two peak values can be evaluated by
$\frac{\partial X(\omega)}{\partial \omega}|_{\omega=\omega_{\pm}}=0$
and $\frac{\partial^{2} X(\omega)}{\partial^{2} \omega}|_{\omega=\omega_{\pm}}<0$,
with $\omega_{\pm}$ the frequencies relevant to the two symmetric peaks.

Based on this feature, we consider below a measurement of the environmental
temperature using one of the peak values (See Fig. 5(c)). The peak value varies linearly
with respect to $T$, in the same fashion as in Fig. 3(b) for the middle peak values in the case
of identical NAMRs. We may find a similar formula to Eq. (11) for the slope of the peak values
$X(\omega_{peak})$ varying with the environmental temperature T as
\begin{eqnarray}
k(\omega_{peak})
=2k_{B}\sum_{j=1}^{2} \gamma_{j}m_{j}[|V_{j}(\omega_{peak}+\omega_{1})|^{2} \nonumber \\
+|V_{j}(-\omega_{peak}+\omega_{1})|^{2}]=k(-\omega_{peak}).
\end{eqnarray}
Straightforward calculations of the slopes in this case
present less precise measurements of the environmental temperature compared
to the case of identical NAMRs since we have the sensitivity with 0.44 ($
\Delta\omega =0.1\omega_{1}$) or 0.33 ($\Delta\omega=0.2\omega_{1}$)
of the counterpart in the case of $\Delta \omega =0$. Therefore, for a more
precise measurement of the environmental temperature with two different
NAMRs, the frequency difference is required to be as tiny as possible.

If the two different NAMRs are with the same frequency but with different
mass, we have only double OMITs, rather than triple OMITs. In this case, we
found that the middle peak remains the same value for different ratios of $%
m_{2}/m_{1}$ provided that the temperature is zero, but varies with
different slopes for different ratios of $m_{2}/m_{1}$ if $T\neq $0 (See
Fig. 6). In particular, for a bigger mass of NAMR$_{2}$ than NAMR$_{1}$,
e.g., $m_{2}=2m_{1}$ in Fig. 6(c), the measurement sensitivity of the
environmental temperature is higher
than the counterpart in the identical case. In addition, the measurement
range changes with different ratios of $m_{2}/m_{1}$ as indicated in Fig.
6(c). Nevertheless, within the range $T\in \lbrack 0,0.12]$ K, we have the
possibility to obtain the measurement precision of temperature better than $%
4.4\times 10^{-5}$ K.

\section{Conclusion}

In summary, we have justified the possibility of precisely detecting the
environmental temperature by the unique quantum characteristics of
double-OMIT. To our knowledge, this is the first scheme for such a precision
measurement in the optomechanical system via the noise spectra. We have
shown by numerical simulation that we are able to have a better precision of
measuring the environmental temperature than a previously reported quantum
thermometer \cite{science-300-1929}.

For simplicity, however, we have remained the radiation pressure as a constant throughout the paper.
For a thorough investigation of the temperature measurement, it is necessary to explore
the change of the radiation pressure. Straightforward calculations indicate that
enhancement of the radiation pressure due to increase of the photon number will definitely lead to a better
precision measurement of the environmental temperature. Nevertheless,
the nonlinear effect in the optomechanics due to more photons involved would bring in unexpected complexity,
which needs further scrutiny.

Moreover, due to the tunable fashion and robustness to cavity decay, the model under
consideration can also be applied to other applications, such as precisely
measuring Coulomb coupling strength and the frequency (mass) difference
between the two NAMRs. Further exploration would be more interesting and is
underway.

\section*{Acknowledgments}

This work is supported by the National Natural Science Foundation of China
under Grants No. 61475045, No. 11274352 and No. 11304366, the China
Postdoctoral Science Foundation (Grants No. 2013M531771 and No.
2014T70760),and Natural Science Funding for Colleges and Universities in
Jiangsu Province (Grant No. 12KJD140002).

\end{document}